\documentstyle[amssymb,aps,preprint,epsf,psfig]{revtex}
\begin{document}
\title{Anomalous self-energy and Fermi surface quasi-splitting in the vicinity of a
ferromagnetic instability}
\author{A. A. Katanin$^{a,b}$, A. P. Kampf$^c,$ and V. Yu. Irkhin$^b$}
\address{$^a$Max-Planck-Institut f\"ur Festk\"orperforschung, 70569 Stuttgart, Germany%
\\
$^b$Institute of Metal Physics, 620219 Ekaterinburg, Russia\\
$^c$Institut f\"ur Physik, Theoretische Physik III,\\
Elektronische Korrelationen und Magnetismus,\\
Universit\"at Augsburg, 86135 Augsburg, Germany}
\maketitle

\begin{abstract}
{We discuss the low-temperature behavior of the electronic self-energy in
the vicinity of a ferromagnetic instability in two dimensions within the
two-particle self-consistent approximation, functional renormalization group
and Ward-identity approaches. Although the long-range magnetic order is
absent at }$T>0,$ the self-energy has a non-Fermi liquid form at low
energies $|\omega |\lesssim \Delta _0$ near the Fermi level, where $\Delta _0
$ is the ground-state spin splitting. The spectral function at temperatures $%
T\lesssim \Delta _0$ has a two-peak structure with finite spectral weight at
the Fermi level. The simultaneous inclusion of self-energy and vertex
corrections shows that the above results remain qualitatively unchanged down
to very low temperatures $T\ll \Delta _0.$ It is argued, that this form of
the spectral functions implies the quasi-splitting of the Fermi surface in
the paramagnetic phase in the presence of strong ferromagnetic fluctuations.
\end{abstract}

\section{Introduction}

Non-Fermi-liquid behavior of correlated low-dimensional electron systems has
attracted much attention during the last decade. This behavior is usually
connected with the violation of the quasiparticle (qp) concept in some
energy window around the Fermi level. A prominent example is the pseudogap
phenomenon observed in underdoped high-T$_c$ compounds \cite{ARPES}. Cuprate
superconductors are however not the only materials example which show a
strong suppression of the low energy spectral weight due to correlation
effects. A qualitatively similar behavior on parts of the Fermi surfaces was
also observed recently in the unconventional superconductor Sr$_2$RuO$_4$ in
an intermediate temperature range\cite{Wang} and the layered manganite
compound La$_{1+x}$Sr$_{2-x}$Mn$_2$O$_7$ \cite{Manganites}.

One possible viewpoint for the origin of the pseudogap in high-T$_c$
compounds is to relate it to precursors of antiferromagnetism \cite
{Kampf,SF1,SF}. The fluctuation exchange (FLEX) \cite{FLEX,FLEX1},
two-particle self-consistent (TPSC) \cite{TPSC}, dynamical cluster
approximation \cite{DCA}, and most recently the functional
renormalization-group (fRG) technique\cite{KK,Rohe} have demonstrated a
strong anisotropy of spectral properties around the Fermi surface (FS) in
the 2D Hubbard model with a possible violation of the qp concept on parts of
the FS. The large incoherent contributions to the electronic spectrum of
low-dimensional metallic antiferromagnets (e.g. an anomalously large
scattering rate) were also discussed in Ref. \cite{IKK1}. These phenomena
result in the formation of a two- \cite{FLEX,FLEX1,TPSC,Rohe} or three-peak
\cite{KK} structure of the spectral function in the vicinity of the
antiferromagnetic (AFM) instability. With decreasing temperature, these
pseudogap features are expected to evolve continuously towards a ground
state spectral function with an AFM energy gap at the Fermi level.

While many results have been obtained for the electronic properties in the
vicinity of an AFM state and some results exist for itinerant ferromagnets
\cite{HertzEdw,IK}, surprisingly much less is known about the evolution of
the quasiparticle properties in the paramagnetic phase near the
ferromagnetic (FM) instability. The paramagnon theory \cite
{Brinkman,DK,Moriya,Stamp} focused mostly on the description of the magnetic
properties, ignoring the renormalization of the one-particle Green function.
Nevertheless, already Doniach and Engelsberg \cite{Doniach} showed that in
three dimensions the qp weight vanishes logarithmically on approaching a FM
zero-temperature quantum phase transition (QPT). For two-dimensional (2D)
systems, an $\varepsilon ^{2/3}$ energy dependence of the self-energy at the
quantum critical point (QCP) can be inferred from similar calculations in
the context of gauge field theories \cite{Lee}, the phase separation problem
\cite{Castellani}, and the Pomeranchuk instability \cite{MetznerSE}. The
latter two instabilities arise also in the zero-momentum transfer
particle-hole channel and therefore are expected to have properties which
are similar to those in the vicinity of the ferromagnetic QPT, at least in
lowest order perturbation theory with respect to the fermion-boson (charge
or spin modes) coupling. The $\varepsilon ^{2/3}$ frequency dependence of
the self-energy implies the vanishing of the qp weight at the Fermi level
and therefore invalidates the qp concept. This may raise doubts about the
validity of the applied scheme, since the above-mentioned calculations did
not consider both, self-energy {\it and} vertex corrections. However, the
calculations by Altshuler et al. \cite{Altshuler} within the gauge field
theory context showed that the $\varepsilon ^{2/3}$ dependence of the
self-energy remains valid also in higher orders of perturbation theory.

The breakdown of the qp concept at the QCP may be even more apparent at
finite temperatures. For fermions interacting with a gauge field it was
shown that for the case of gapless (although diffusive) bosonic excitations
the imaginary part of the self-energy in a non-self-consistent calculation
is divergent at the Fermi level at $T>0$ as a consequence of the diverging
static spin susceptibility $\chi ({\bf 0},0)$ \cite{Lee}. This divergence
should necessarily have certain consequences for the zero-momentum
particle-hole instabilities of fermion systems with short-range
interactions. Although the magnetic correlation length $\xi $ of 2D systems
with short-range interaction is finite at finite $T$ and $\chi ({\bf 0},0)$ $%
\propto \xi ^2$, three temperature regimes should be distinguished \cite
{Sachdev} (see Fig. 1): (i) the quantum disordered regime with a disordered
ground state and almost temperature-independent correlation length, (ii) the
quantum critical (QC) regime with $\xi \sim 1/T,$ and (iii) the
renormalized-classical (RC) regime above an ordered ground state with an
exponentially large correlation length at low $T$. The divergence of the
imaginary part of the self-energy for $\xi \rightarrow \infty $ may lead to
especially strong effects in the RC regime, where the inverse correlation
length is almost negligibly small at low $T.$ Indeed, for the AFM
instability this divergence results in the formation of a pseudogap
structure in the spectral function\cite{TPSC}. Although similar properties
in the vicinity of a FM instability were discussed quite recently \cite
{TPSC1,Month}, the behavior in this case is far less from clear, since the
suppression of the spectral weight at the Fermi level itself weakens the
tendency to ferromagnetic order, which therefore implies the necessity to
account for self-energy and vertex corrections selfconsistently.

The recently found non-analytic contributions to the spin susceptibility in
second-order perturbation theory with respect to the electron-electron
interaction \cite{Nonan} have led to infer the possible absence of a
second-order FM QPT, which is either replaced by a first-order transition to
ferromagnetism or a second-order QPT into an incommensurate phase with a
finite ordering wavevector ${\bf Q}$, which then continuously decreases
towards ${\bf Q=0}$ on moving away from the QPT into the ordered state.
However, these corrections are not expected, at least in the weak-coupling
regime, to remove the renormalized classical temperature regime entirely.
Indeed, the corresponding characteristic temperature scale $T_X$ $\sim $($%
U/4\pi t$)$^2t,$ below which these corrections become important\cite{Nonan},
is quadratic in the interaction and therefore small in the weak-coupling
regime in comparison with both the bandwidth and the crossover temperature $%
T^{*}\sim U$ into the renormalized classical regime not too close to the QPT$%
.$

It therefore appears demanding to investigate the finite-temperature
behavior of the self-energy in the vicinity of the FM instability and to
compare non-self-consistent and self-consistent techniques. For
non-self-consistent calculations, the TPSC as well as the recently proposed
fRG approaches on a patched FS \cite{Zanchi,Metzner,SalmHon,SalmHon1} can be
used. The latter approaches have the advantage that they do not select
certain types of electronic scattering processes, but consider them all on
equal footing. The self-consistent treatment of self-energy and vertex
corrections is per se a rather difficult task. Most consistently this can be
done in a parquet-type analysis\cite{Parq} (see also Refs. \cite
{Bickers1,JanisR}), which is however necessarily rather involved for
numerical studies in dimensions $d>1$. In the fRG technique the back
influence of the self-energy on one- and two-particle properties requires to
work at two-loop order, which is currently inachievable. To obtain
qualitative results deep inside the RC regime, the application of Ward
identities offers an alternative. This strategy was chosen previously by
Edwards and Hertz \cite{HertzEdw} to calculate the spin-resolved self-energy
in the ordered FM phase. As we will show in this paper, an analogous
approach can be similarly applied for magnetically disordered systems in the
renormalized-classical regime.

In the present paper we use the TPSC,\ fRG, and Ward identity approaches to
get insight from three different points of view into the behavior of the
self-energy in 2D systems at finite temperatures on approaching the FM
phase. First, in Sect. 2 we consider the main features of the self-energy
obtained within the TPSC approximation and compare them with the results of
the fRG approach, which is used to study in detail the frequency dependence
of the self-energy at van Hove (vH) band fillings. In Sect. 3 we make use of
Ward identities for a qualitative insight into the role of self-energy and
vertex corrections at low temperatures.

\section{The self-energy in non-self-consistent approaches}

Specifically we consider the Hubbard model for $N_e$ electrons on a square
lattice,
\begin{equation}
H=-\sum_{ij\sigma }t_{ij}c_{i\sigma }^{\dagger }c_{j\sigma
}+U\sum_in_{i\uparrow }n_{i\downarrow }-(\mu -4t^{\prime })N_e  \label{H}
\end{equation}
where the hopping amplitude is $t_{ij}=t$ for nearest-neighbor sites $i$ and
$j$ and $t_{ij}=-t^{\prime }$ for next-nearest-neighbor sites ($t,t^{\prime
}>0$); for further convenience we have shifted the chemical potential $\mu $
by $4t^{\prime }$; the corresponding electronic dispersion is
\begin{equation}
\varepsilon _{{\bf k}}=-2t(\cos k_x+\cos k_y)+4t^{\prime }(\cos k_x\cos
k_y+1)-\mu .
\end{equation}
Furthermore we compare some of our results for the Hubbard model with those
for the ferromagnetic s-d model \cite{Vonsovskii,Nagaev1}
\begin{equation}
H=-\sum_{ij\sigma }t_{ij}c_{i\sigma }^{\dagger }c_{j\sigma }-I\sum_i{\bf S}%
_i \cdot\mbox{\boldmath $\sigma $}_{\sigma \sigma ^{\prime }}c_{i\sigma
}^{\dagger }c_{i\sigma ^{\prime }}-\frac 12\sum_{ij}J_{ij}{\bf S}_i\cdot
{\bf S}_j-(\mu -4t^{\prime })N_e  \label{sd}
\end{equation}
in the weak-coupling regime $0<I\ll 8t,$ where ${\bf S}_i$ are
localized-spin operators and $\mbox{\boldmath $\sigma
$}_{\sigma \sigma ^{\prime }}$ are Pauli matrices; $J_{ij}=J>0$ is the
direct nearest-neighbor ferromagnetic spin exchange coupling.

\subsection{TPSC approximation}

The TPSC approximation\cite{TPSC} for the Hubbard model (\ref{H}) considers
the dynamical spin susceptibility
\begin{equation}
\chi ({\bf q},\omega )=\frac{\chi _0({\bf q},\omega )}{1-U_{\text{sp}}\chi
_0({\bf q},\omega )}  \label{Hi}
\end{equation}
which has the same structure as the spin susceptibility in the random-phase
approximation (RPA), but contains an effective interaction $U_{\text{sp}}$
instead of the bare $U.$ The bare susceptibility $\chi _0({\bf q},i\omega
_n) $ is given by
\begin{equation}
\chi _0({\bf q},i\omega _n)=\frac 1N\sum_{{\bf k}}\frac{f_{{\bf k}}-f_{{\bf %
k+q}}}{i\omega _n-\varepsilon _{{\bf k}}+\varepsilon _{{\bf k+q}}}
\end{equation}
where $f_{{\bf k}}=f(\varepsilon _{{\bf k}})$ is the Fermi function and $%
\omega _n=2\pi nT$. The effective interaction $U_{\text{sp}}$ is determined
by the sum rule \cite{TPSC}
\begin{equation}
\frac TN\sum_{{\bf q,}i\omega _n}\chi ({\bf q},i\omega _n)=n/2-n^2U_{\text{sp%
}}/(4U)  \label{sr}
\end{equation}
with $n=N_e/N$ being the band filling and $k_B$ is set to unity. The
renormalization of the interaction $U_{sp}/U<1$ avoids the artificial
divergence of the susceptibility at finite temperature where the denominator
of Eq. (\ref{Hi}) calculated with the bare interaction $U$ vanishes. The
temperature and $t^{\prime }$ dependence of $U_{\text{sp}}$ was extensively
discussed in Refs. \cite{TPSC,TPSC1}; $U_{sp}$ decreases with increasing $%
t^{\prime }$ and decreasing temperature.

Above the ordered ground state the inverse correlation length in TPSC
\begin{equation}
\xi ^{-1}\propto [1-U_{\text{sp}}\chi _0({\bf Q},0)]^{1/2}  \label{ksi}
\end{equation}
monotonically decreases with temperature and becomes exponentially small, $%
\xi ^{-1}=C\exp (-T^{*}/T)$, below a certain crossover temperature $%
T^{*}=4\pi AU_{\text{sp}}\overline{S}_0^2/(3\overline{\chi }_0),$ where $%
\overline{\chi }_0=\chi _0({\bf Q},0),$ $A=\nabla ^2\chi _0({\bf q},0)|_{%
{\bf q}={\bf Q}},$ $C$ is a constant, and
\begin{equation}
\overline{S}_0^2=3n/4-3n^2/(8U\overline{\chi }_0)-\frac 32\left[ \frac TN%
\sum_{{\bf q,}i\omega _n}\frac{\chi _0({\bf q},i\omega _n)}{1-\chi _0({\bf q}%
,i\omega _n)/\overline{\chi }_0}\right] _{T=0}
\end{equation}
is the square of the ground-state (sublattice) magnetization per site, ${\bf %
Q}$ is the magnetic ordering wavevector, determined by the maximum of $\chi
_0({\bf Q},0).$ In the following we suppose ${\bf Q=0}${\bf ,} which
corresponds to a ferromagnetic instability and is in particular the case for
vH band fillings ($\mu =0$) of the $t$-$t^{\prime }$ Hubbard model, Eq. (\ref
{H}), with $0.3t\lesssim t^{\prime }<0.5t$ \cite{SalmHon1,KK1,TPSC1}. Note
that we ignore here the possibility of triplet pairing, which may also arise
in the vicinity of the FM instability\cite{SalmHon1}.

The self-energy is given by
\begin{equation}
\Sigma ({\bf k},i\varepsilon _n)=\frac{UU_{\text{sp}}T}{2N}\sum_{{\bf q,}
i\omega _n}[3\chi ({\bf q},i\omega _n)-\chi _0({\bf q},i\omega _n)]\frac 1{
i\varepsilon _n+i\omega _n-\varepsilon _{{\bf k+q}}}  \label{SS}
\end{equation}
on the imaginary frequency axis [$\varepsilon _n=(2n+1)\pi T$ are fermionic
Matsubara frequencies] and
\begin{equation}
\Sigma ({\bf k},\varepsilon +i0^{+})=\frac{UU_{\text{sp}}}{2N}\sum_{{\bf q}
}\int \text{d}\omega [3\text{Im}\chi ({\bf q},\omega )-\text{Im}\chi _0({\bf %
q},\omega )]\frac{N_B(\omega )+f(\varepsilon _{{\bf k+q}})}{\varepsilon
+\omega -\varepsilon _{{\bf k+q}}+i0^{+}}  \label{SS1}
\end{equation}
on the real axis, where $N_B(\omega )$ is the Bose distribution function.
The factor $3$ in the first term in the square brackets of Eqs. (\ref{SS})
and (\ref{SS1}) arises from the summation over three (2 transverse and 1
longitudinal) spin channels, the second term in the square brackets is
substracted to avoid double counting of the second-order diagram.

The same expressions (\ref{SS}) and (\ref{SS1}) with the replacement $UU_{%
\text{sp}}\rightarrow I^2$ hold for the s-d model (\ref{sd}) in second order
perturbation theory with respect to the electron-spin interaction $I$, see,
e.g. Ref. \cite{IK}. The magnetic correlations in this case originate mainly
from the exchange interaction between the localized spins, and the magnetic
susceptibility has the same form as for the Heisenberg model with nearest
neighbor exchange interaction $J$. The inverse magnetic correlation length
at low $T$ is again exponentially small \cite{Chakravarty}, $\xi ^{-1}=C\exp
(-JS^2/T);$ the corresponding crossover temperature into the RC regime is $%
T^{*}\sim JS^2.$

To calculate the self-energy at small $\varepsilon $ and ${\bf k}$ near the
Fermi surface, we expand the bare susceptibility $\chi _0$ at small $q$ and $%
\omega .$ For the Hubbard model at $\mu \neq 0$ the resulting spin
susceptibility on the real frequency axis is given by (see e.g. Ref. \cite
{Moriya})
\begin{equation}
\chi ({\bf q},\omega )=\frac{\chi _0}{A(q^2+\xi ^{-2})+iB\omega /q},
\label{Hi1}
\end{equation}
where the correlation length $\xi $ in TPSC is given by Eq. (\ref{ksi}), $A$
and $B$ are constants which are proportional to $U_{\text{sp}}\ $with a
coefficient which depends on the bare spectrum $\varepsilon _{{\bf k}}$ (see
Ref. \cite{Moriya} for explicit expressions). For the van Hove singularity
case ($\mu =0$) we obtain
\begin{equation}
\chi ({\bf q},\omega )=\frac{\chi _0}{A(q^2+\xi ^{-2})+i\Gamma (q_{\pm
},\omega )}  \label{Hi2}
\end{equation}
where (cf. the $T=$ $t^{\prime }=0$ result of Ref. \cite{Guinea})
\begin{equation}
\Gamma (q_{\pm },\omega )=\frac{U_{\text{sp}}}{2\pi t\sin ^22\varphi }\frac
\omega {\max [\omega ,(Tt|q_{+}q_{-}|)^{1/2},t|q_{+}q_{-}|]}
\end{equation}
is the damping of the spin excitations, $q_{\pm }=q_x\sin \varphi \pm
q_y\cos \varphi ,$ and $\cos 2\varphi =2t^{\prime }/t$.

The form of the dynamic magnetic susceptibility of the s-d model in the
paramagnetic phase, which is mostly determined by the local moment
subsystem, is more complicated. However, we will see below that already the
static part of the susceptibility dominates the spectral properties at low $%
T\lesssim T^{*}$. In the static limit the susceptibility in the RC regime is
expected to coincide with that determined for the AFM case within a $1/M$
expansion in the $O(M)$ Heisenberg model \cite{Chubukov} and has the same
form as Eq. (\ref{Hi1}) with $\omega =0.$

To analyze spectral properties, we first consider the inverse qp lifetime
\begin{equation}
\frac 1{\tau ({\bf k}_F)}=-\text{Im}\Sigma ({\bf k}_F,{\rm i}0^+)=-\frac{\pi
UU_{\text{sp}}}2N\sum_{{\bf q}}\frac{3\text{Im}\chi ({\bf q},\varepsilon _{%
{\bf k}_F{\bf +q}})-\text{Im}\chi _0({\bf q},\varepsilon _{{\bf k}_F{\bf +q}%
})}{ \sinh (\varepsilon _{{\bf k}_F{\bf +q}}/T)}  \label{ImS}
\end{equation}
and the derivative of the real part of the self-energy at the Fermi level
\begin{equation}
\left. \frac{\partial \text{Re}\Sigma ({\bf k}_F,\varepsilon )}{\partial
\varepsilon }\right| _{\varepsilon =0}=-\frac{UU_{\text{sp}}}{2N}{\cal P}%
\int {\rm d}\omega \sum_{{\bf q}}[3\text{Im}\chi ({\bf q},\omega )-\text{Im}%
\chi _0({\bf q},\omega )]\frac{N_B(\omega )+f(\varepsilon _{{\bf k}_F{\bf +q}%
})}{(\omega -\varepsilon _{{\bf k}_F{\bf +q}})^2}.
\end{equation}
The results for the leading terms at $T\ll t$ together with the results of
second-order perturbation theory (SOPT, which is obtained by the replacement
$\chi \rightarrow \chi _0$ in Eq. (\ref{ImS})) are collected in Table I,
where we omit overall temperature-independent prefactors which are
proportional to $U^2/t^2$.
\[
\begin{tabular}{||l|ccc|c||}
\hline\hline
Spectrum & \multicolumn{2}{|c|}{$\tau ^{-1}({\bf k}_F)$} &
\multicolumn{2}{|c||}{$[\partial $Re$\Sigma ({\bf k}_F,\varepsilon
)/\partial \varepsilon ]_{\varepsilon =0}$} \\ \hline\hline
& \multicolumn{1}{|c|}{SOPT} & \multicolumn{1}{c|}{TPSC} & SOPT &
\multicolumn{1}{c||}{TPSC} \\ \hline
Linear ($\mu \neq 0$) & \multicolumn{1}{|c|}{$(T^2/t)\ln (1/T)$} &
\multicolumn{1}{c|}{$T\xi +t{\cal O}((T/t)^{2/3})$} & const$\,<0$ &
\multicolumn{1}{c||}{$(T/t)\xi ^2+{\cal O}((T/t)^{1/3})$} \\ \hline
vHs ($\mu =0$)$\left\{
\begin{array}{c}
{\bf k}={\bf k}_{\text{vH}} \\
{\bf k}\neq {\bf k}_{\text{vH}}
\end{array}
\right. $ & \multicolumn{1}{|l|}{$
\begin{array}{c}
T\ln (1/T) \\
(T^2/t)\ln (1/T)
\end{array}
$} & \multicolumn{1}{l|}{$
\begin{array}{c}
T\xi ^2+t{\cal O}((T/t)^{2/3}) \\
T\xi +t{\cal O}((T/t)^{5/6})
\end{array}
$} & \multicolumn{1}{l|}{$
\begin{array}{c}
\ln (1/T) \\
\text{const}<0
\end{array}
$} & \multicolumn{1}{|l||}{$
\begin{array}{c}
(T/t)\xi ^4+{\cal O}((T/t)^{1/3}) \\
(T/t)\xi ^2+{\cal O}((T/t)^{2/3})
\end{array}
$} \\ \hline\hline
\end{tabular}
\]

Table I. Inverse quasiparticle lifetime $\tau ^{-1}({\bf k}_F)$ and $%
[\partial $Re$\Sigma ({\bf k}_F,\varepsilon )/\partial \varepsilon
]_{\varepsilon =0}$ in second-order perturbation theory (SOPT) and the
two-particle self-consistent approximation (TPSC) near the FM instability
for different bare electronic spectra.

\vskip0.5cm Apparently, the imaginary part of the self-energy at the Fermi
level is anomalously enhanced by the correlation effects for $\xi \gg 1$ and
even tends to diverge deep inside the RC regime $T\ll T^{*}$ where $\xi
\rightarrow \infty .$ Simultaneously, $\partial $Re$\Sigma /\partial
\varepsilon |_{\varepsilon=0}$ becomes positive and large. At energies $%
|\omega |\gtrsim t\xi^{-1}$ ($t\xi ^{-2}$ in the vHs case) the real part of
the self-energy behaves as $\Delta _0^2/\omega ,$ where $\Delta
_0=(UU_{sp})^{ 1/2}\overline{S}_0\sim T^{*}$. It is worthwhile to note that
the abovementioned divergencies arise from the purely static contributions
with zero bosonic Matsubara frequency and were previously discussed in
detail for the AFM case in Refs. \cite{SF,TPSC}, where $\overline{S}_0$ is
the ground-state sublattice magnetization.

These low-energy features lead to a suppression of spectral weight in $A(
{\bf k},\omega)=-(1/\pi)$Im$\Sigma /[(\omega -\varepsilon _{{\bf k}}+{\rm Re}%
\Sigma )^2+({\rm Im}\Sigma )^2]$ at $|\omega |\lesssim $ $\Delta _0$ (see
also Ref. \cite{TPSC1}) and to the formation of a two-peak structure of the
spectral function with the peaks located near $\omega \simeq \pm \Delta _0$.
While for the AFM case the fulfillment of the nesting condition $\varepsilon
_{{\bf k}_F+{\bf Q}}=-\varepsilon _{{\bf k}_F}=0$ (which for $t^{\prime
}\neq 0$ is satisfied only at the hot spots) is required, in the FM case the
two-peak structure occurs all around the FS. The difference between the FM
and AFM instabilities at those points where $\varepsilon _{{\bf k}_F+{\bf Q}%
}=0$ is evident only in the subleading terms $\sim T^\alpha $, where the
exponent $\alpha$ depends on the dynamical exponent $z$ [e.g. $\tau ^{-1}(%
{\bf k}_F) \sim T\xi+t(T/t)^{1-1/z}$ and $\partial $Re$\Sigma /\partial
\varepsilon |_{\varepsilon =0}\sim (T/t)\xi ^2+(T/t)^{1-2/z}$ for the linear
electronic dispersion near the Fermi level], $z=2$ for the AFM and $z=3$ for
the FM case. Note that unlike Refs. \cite{TPSC,TPSC1} we did not suppose
that $T\ll T^{*}$ in deriving the results of Table I, and in fact these
results are valid for both, RC and QC regimes.

The above discussed features of the self-energy keep their form away from
the FS with the replacement $\varepsilon \rightarrow \varepsilon
-\varepsilon _{{\bf k+Q}}.$ This holds for all $|\varepsilon _{{\bf k+Q}%
}|\ll t$ for a linear electronic dispersion and for $t\xi ^{-2}\ll
|\varepsilon _{{\bf k+Q}}|\ll t$ for vH band fillings. At vH band fillings
and $|\varepsilon _{{\bf k+Q}}|\lesssim t\xi ^{-2}$ additional divergent
terms in the real part of the self-energy arise at the vH points ${\bf k}_{%
\text{vH}}=$($0,\pi $) and ($\pi ,0$) with Re$\Sigma ({\bf k}_{\text{vH}%
},0)\sim tk_{+}k_{-}\xi ^4$ [$k_{\pm }=(k_x-\pi )\sin \varphi \pm k_y\cos
\varphi $ for ${\bf k}_{\text{vH}}=$($\pi ,0$) and similarly for the other
vH point], which flatten the bare electronic dispersion at the momenta near $%
{\bf k}_{\text{vH}}$, similar to the zero temperature case \cite{Pinning}.
This flattening however is not important in the RC regime since the
corresponding momentum region is rather narrow and Im$\Sigma $ at the Fermi
level is finite at finite $T$ (and even diverges at $\xi \rightarrow \infty $%
).

The dependence on $\varepsilon -\varepsilon _{{\bf k+Q}}$ for $t\xi ^{-2}\ll
|\varepsilon _{{\bf k+Q}}|\ll t$ is the origin for an important difference
between the spectral functions near the FM and AFM instabilities {\it away}
from the FS. In the FM case the spectral functions depend on $\varepsilon
-\varepsilon _{{\bf k}}$ only, implying that at $\varepsilon _{{\bf k}%
}\simeq \pm \Delta _0$ (the condition which determines the FSs of spin-up
and spin down electrons in the FM phase) one of the peaks of the above
discussed two-peak structure is located at the Fermi level, making the
electronic excitations at the points of the Brillouin zone with $\varepsilon
_{{\bf k}}\simeq \pm \Delta _0$ coherent. This indicates the existence of
two ``preformed'' Fermi surfaces already in the paramagnetic (PM) phase at
low temperatures $T\lesssim T^{*}$. The corresponding electronic
excitations, however, do not have any prefered spin direction and the spin
symmetry remains necessarily unbroken at $T>0.$ As we show in Section III,
self-consistent approaches show the same tendency of the spectral weight
suppression at the PM Fermi surface, and the redistribution of spectral
weight towards the energies $\omega \simeq \pm \Delta _0$ supports therefore
the picture described above in the non-self-consistent TPSC analysis.

\subsection{fRG analysis at van Hove band fillings}

The two-particle self-consistent approximation may be insufficient close to
van Hove band fillings since it considers only the contribution of
particle-hole excitations and the other electronic scattering channels are
accounted for only by the $``$average'' renormalized interaction $U_{sp}.$
Moreover, at van Hove band fillings the effective interaction $U_{sp}$
artificially tends to zero with decreasing temperature (see e.g. Ref. \cite
{TPSC}). To analyze in more detail the frequency dependence of the
self-energy in the vicinity of van Hove band fillings, we apply the fRG
approach for one particle-irreducible (1PI) functions \cite{SalmHon} with a
temperature cutoff \cite{SalmHon1}. This approach considers the evolution of
the generating functional with decreasing temperature in the weak-coupling
regime. The flow of the self-energy $\Sigma _T({\bf k},i\omega
)=T^{-1/2}\Sigma ({\bf k},i\omega )$ in the 1PI fRG scheme is given by
\begin{equation}
\frac{{\rm d}\Sigma _T}{{\rm d}T}=V_T\circ S_T,  \label{OneLoopSE}
\end{equation}
where $\circ $ is a short notation for the summation over momentum,
frequency and spin variables according to standard diagrammatic rules, see
e.g. Ref. \cite{SalmHon2}. The renormalization of the electron-electron
interaction vertex $V_T$ at one-loop order is given by
\begin{equation}
\frac{{\rm d}V_T}{{\rm d}T}=V_T\circ (G_T\circ S_T+S_T\circ G_T)\circ V_T.
\label{OneLoop}
\end{equation}
The propagators $G_T$ and $S_T\ $are defined by
\begin{equation}
G_T({\bf k},i\nu _n)=\frac{T^{1/2}}{i\nu _n-\varepsilon _{{\bf k}}};\,\;S_T(%
{\bf k},i\nu _n)=-\frac 1{2T^{1/2}}\frac{i\nu _n+\varepsilon _{{\bf k}}}{%
(i\nu _n-\varepsilon _{{\bf k}})^2}.  \label{GS}
\end{equation}
The factors $T^{1/2}$ arise due to the rescaling of the fermion fields on
removing the temperature dependence from the interaction term of the action
\cite{SalmHon1}. Eqs. (\ref{OneLoopSE}) and (\ref{OneLoop})\ have to be
solved with the initial conditions $V_{T_0}=U$ and $\Sigma _{T_0}=0$ where $%
T_0\gg t.$ In the non-self-consistent treatment in this Section we have
neglected the self-energy in the denominators of the Green functions (\ref
{GS}). The self-consistent RG analysis is rather involved and requires the
inclusion of two-loop corrections which remain a challenging task for fRG
techniques.

Since the frequency dependence of the vertices is neglected in the
calculations, it is convenient to reinsert, following Ref. \cite{SalmHon2},
the vertex from Eq. (\ref{OneLoop}) into Eq. (\ref{OneLoopSE}) to obtain
\begin{equation}
\frac{{\rm d}\Sigma _T}{{\rm d}T}=S_T\circ \int\limits_T^{T_0}{\rm d}
T^{\prime }[V_{T^{\prime }}\circ (G_{T^{\prime }}\circ S_{T^{\prime
}}+S_{T^{\prime }}\circ G_{T^{\prime }})\circ V_{T^{\prime }}].  \label{Se2}
\end{equation}
To avoid the integration over temperature in the right-hand side of Eq. (\ref
{Se2}) we integrate by parts to obtain (cf. Ref. \cite{SalmHon2} for the
momentum-cutoff scheme)
\begin{equation}
\frac{{\rm d}\Sigma _T}{{\rm d}T}=(G_T-G_{T_f})\circ [V_T\circ (G_T\circ
S_T+S_T\circ G_T)\circ V_T]  \label{Se21}
\end{equation}
where $T_f$ is the final temperature where the self-energy is evaluated.
Although Eq. (\ref{Se21}) contains the Matsubara sums of the Green functions
at different temperatures, these sums can be calculated by the same
procedure as for equal-temperature Green functions.

To solve Eqs. (\ref{OneLoop}) and (\ref{Se21}) numerically we divide the
momentum space into 32 patches with the same patching scheme as in Refs.\cite
{SalmHon,SalmHon1}. To calculate the self-energy on the real frequency axis
we use analytical continuation by Pad\'e approximants \cite{Pade}. Similar
to Ref. \cite{KK} we use the advantage of Eq. (\ref{Se2}) that for
frequency-independent vertices, after analytical summation over internal
frequencies, the self-energy can be calculated at arbitrary frequencies on
the imaginary axis, and therefore we choose a mesh on the frequency axis,
which becomes denser close to $i\omega =0$ \cite{KK}.

We consider the results at the vH band filling ($\mu =0$) for $t^{\prime
}=0.45t$ and $U=4t$ where ferromagnetism is expected in the ground state\cite
{SalmHon1,KK1}. We choose this relatively large value of $U$ because the
crossover temperatures $T^{*}$ for the FM instability are smaller then for
the AFM case ($T^{*}\lesssim 0.1t$ for the parameters we use), and for lower
values of $U$ and correspondingly lower temperatures the analytical
continuation becomes increasingly difficult, since the size of the anomalous
frequency region in the vicinity of the Fermi level decreases with $T^{*}$.
The self-energy in second-order perturbation theory (SOPT, which is obtained
by the replacement $V\rightarrow U$ in Eq. (\ref{Se2})) calculated at
temperatures $T=0.1t,0.3t,$ and $0.5t$ for ${\bf k}_F=(2.83,0.07),$ which is
the center of the first patch, closest to the $(\pi ,0)$ point, is shown in
Fig. 2. The SOPT self-energy at other points on the FS looks similar. As in
the AFM case \cite{KK} at high temperatures $T\gtrsim 0.3t$ the self-energy
has a sharp dip at the Fermi level due to vH singularity effects. Im$\Sigma
^{(\text{SOPT})}({\bf k}_F,0)$ decreases with decreasing temperature and Im$%
\Sigma ({\bf k}_F,0)=0$ at $T=0.$ This however does not imply the validity
of the qp picture everywhere on the Fermi surface since Im$\Sigma ({\bf k}%
_F,\varepsilon )\propto \varepsilon \ln (1/\varepsilon )$ for the vH points $%
{\bf k}_F=(\pi ,0)$ and $(0,\pi )$ at $T=0$\cite{Guinea}, although the
``normal'' 2D behavior Im$\Sigma ({\bf k}_F,\varepsilon )\propto \varepsilon
^2\ln (1/\varepsilon )$ \cite{Bloom} is restored for other points of the
Fermi surface.

The self-energy obtained within the fRG technique for the temperatures $%
T=0.1t$ and $T=0.3t$ is shown in Figs. 3-5. The results for $T=0.5t$ (not
shown) are similar to those at $T=0.3t$ and both are close to the results of
the SOPT, Fig. 2. But in contrast to SOPT, with decreasing temperature to $%
T=0.1t$ a new sharp feature appears in a narrow region near the Fermi level
(Fig. 4). Note that the maximum effective interaction $V_{\max }\equiv \max
\{V({\bf k}_1,{\bf k}_2;{\bf k}_3,{\bf k}_4)\}=20t$ for this temperature$.$
Similarly to the TPSC approach of Sect. IIa, the imaginary part of the
self-energy has a minimum at the Fermi level instead of a maximum as
expected for a Fermi liquid. Simultaneously, Re$\Sigma ({\bf k}_F,\omega )$
has a positive slope near $\omega =0$. These pronounced anomalies in the fRG
self-energy increase in size with decreasing $T$ and lead to a suppression
of spectral weight at the Fermi energy. The spectral function (Fig. 4d) has
an asymmetric two-peak structure. A qualitatively similar picture is
observed in the fourth FS patch, closest to the Brillouin zone diagonal with
${\bf k}_F=(1.06,0.75)$ (Fig. 5) where at low temperatures the fRG result
also leads to a two-peak structure of the spectral function, where however
the $\omega >0$ peak is larger than $\omega <0$ one.

The magnitude of Re$\Sigma ({\bf k}_F,0)\simeq -0.4t$ in the first, closest
to $(\pi,0)$ patch of the FS is almost the same in both, SOPT and fRG
approaches, and is much larger than the corresponding value found in the
vicinity of the AFM instability \cite{KK}. At the same time, in the fourth
patch of the FS, the fRG approach gives slightly positive Re$\Sigma ( {\bf k}%
_F,0)$, while Re$\Sigma ({\bf k}_F,0)$ in SOPT approach remains almost
patch-independent. Therefore, the fRG approach, in contrast to SOPT, leads
to a significant deformation of the FS near the FM instability at vH band
fillings, which reflects a band narrowing tendency; in essence, the
effective (renormalized) value of the next-nearest neighbor hopping $%
t^{\prime}$ increases towards its value in the flat-band case, $t^{\prime
}/t=1/2$ (see e.g. the discussion in Ref. \cite{KK1}). This is similar to
the earlier discussed flattening of the dispersion close to ($\pi ,0$) and ($%
0,\pi$) points at vH band fillings\cite{Pinning}, except that in the present
case the flattening affects larger parts of the Brillouin zone along the
zone axes. The sizable value of $|$Re$\Sigma |$ at the Fermi surface leads
to an asymmetric shape of the spectral functions in the vicinity of the
Fermi energy, in contrast to the analytical approach of Sect IIa, where this
FS shift was not taken into account.

\section{The self-consistent approach at $T\ll T^{*}$}

To get qualitative insight into the role of self-consistency effects below $%
T^{*}$, we consider the general form of the spin susceptibilities
\begin{eqnarray}
\chi ^{+-}({\bf q},i\omega _n) &=&-\frac TN\sum_{{\bf k},i\nu _n}\Gamma
_{\uparrow \downarrow }^{+}({\bf k},{\bf k+q;}i\nu _n,i\nu _n+i\omega
_n)G_{\uparrow }({\bf k},i\nu _n)G_{\downarrow }({\bf k+q},i\omega _n+i\nu
_n)  \nonumber \\
\chi ^{zz}({\bf q},i\omega _n) &=&-\frac TN\sum_{{\bf k},i\nu _n}\Gamma
_{\uparrow \uparrow }^z({\bf k},{\bf k+q;}i\nu _n,i\nu _n+i\omega
_n)G_{\uparrow }({\bf k},i\nu _n)G_{\uparrow }({\bf k+q},i\omega _n+i\nu _n)
\label{hig}
\end{eqnarray}
where
\begin{eqnarray}
\Gamma _{\sigma \sigma ^{\prime }}^a({\bf k},{\bf k}^{\prime },i\nu _n,i\nu
_n^{\prime }) &=&G_\sigma ^{-1}({\bf k},i\nu _n)G_{\sigma ^{\prime }}^{-1}(%
{\bf k}^{\prime },i\nu _n^{\prime })  \nonumber \\
&&\ \ \times \int\limits_0^{1/T}\int\limits_0^{1/T}e^{i(\nu _n\tau +\nu
_n^{\prime }\tau ^{\prime })}\langle T[S_{{\bf k-k}^{\prime }}^a(0)c_{{\bf k}%
\sigma }^{\dagger }(\tau )c_{{\bf k}^{\prime },\sigma ^{\prime }}(\tau
^{\prime })]\rangle d\tau d\tau ^{\prime }  \label{SEW}
\end{eqnarray}
are the 3-point spin vertex functions and $S_{{\bf q}}^a(\tau )=\frac 1N%
\sum_{{\bf k}}c_{{\bf k}\alpha }^{\dagger }(\tau )\sigma _{\alpha \beta
}^ac_{{\bf k+q},\beta }(\tau )$. We retain the spin indices in this section
to perform calculations in the presence of a small static external Zeeman
field $h$ which is set to zero at the end. To arrive at the RPA-like form of
the susceptibilities, we follow Ref. \cite{HertzEdw} and introduce the
irreducible vertices $\gamma _{\sigma \sigma ^{\prime }}^a$%
\begin{equation}
\gamma _{\sigma \sigma ^{\prime }}^a({\bf k},{\bf k}^{\prime },i\nu _n,i\nu
_n^{\prime })=\frac{\Gamma _{\sigma \sigma ^{\prime }}^a({\bf k},{\bf k}%
^{\prime },i\nu _n,i\nu _n^{\prime })}{1+U\chi _{\sigma ,\sigma ^{\prime
}}^a({\bf k}-{\bf k}^{\prime },i\nu _n-i\nu _n^{\prime })}  \label{Gg}
\end{equation}
to obtain
\begin{equation}
\chi ^{ab}({\bf q},i\omega _n)=\frac{\phi ^{ab}({\bf q},i\omega _n)}{1-U\phi
^{ab}({\bf q},i\omega _n)}  \label{hifi}
\end{equation}
where $a,b=+,-,z,$ and the function $\phi ^{ab}$ has the same structure as $%
\chi ^{ab}$ in Eq. (\ref{hig}) with the replacement $\Gamma \rightarrow
\gamma .$ Note that
\[
\gamma _{\uparrow \downarrow }^{+}=\gamma _{\downarrow \uparrow }^{-}=\gamma
_{\uparrow \uparrow }^z=-\gamma _{\downarrow \downarrow }^z;\chi ^{+-}=2\chi
^{zz}
\]
(and similar for $\Gamma $) at $h=0.$

The expression for the self-energy, including the Hartree term, reads
\begin{eqnarray}
\Sigma _\sigma ({\bf k},i\nu _n) &=&-\frac{TU}N\sum_{{\bf k}^{\prime
},iv_n^{\prime }}G_\sigma ({\bf k}^{\prime },i\nu _n^{\prime })  \nonumber \\
&&+\frac{TU^2}N\sum_{{\bf q},i\omega _n}[\gamma _{\sigma ,\sigma }^z({\bf k},%
{\bf k}+{\bf q};i\nu _n,i\nu _n+i\omega _n)G_\sigma ({\bf k}+{\bf q},i\omega
_n+i\nu _n)\chi ^{zz}({\bf q},i\omega _n)  \nonumber \\
&&+\gamma _{\sigma ,-\sigma }^{+}({\bf k},{\bf k}+{\bf q};i\nu _n,i\nu
_n+i\omega _n)G_{-\sigma }({\bf k}+{\bf q},i\omega _n+i\nu _n)\chi ^{+-}(%
{\bf q},i\omega _n)]  \label{Si}
\end{eqnarray}
We neglected here the contribution of charge excitations which are not
singular near the magnetic phase transition.

We note that a similar expression for the self-energy can be obtained for
the ferromagnetic s-d model (\ref{sd}) but in this case the susceptibilities
$\chi ^{zz}({\bf q},i\omega _n)$ and $\chi ^{+-}({\bf q},i\omega _n)$ are
determined by the localized-moment subsystem. This difference however will
not be important in the following sequence of arguments and the results of
this section are applicable for this model as well.

As discussed in Sect. IIa, the main contribution to Eq. (\ref{SEW}) comes
from the bosonic Matsubara frequency ${\rm i}\omega _n=0.$ The vertex $%
\Gamma _{\sigma ,\sigma ^{\prime }}^a({\bf k},{\bf k+q};i\nu _n,i\nu _n)$ in
the limit ${\bf q\rightarrow 0}$ [which we denote in the following as $%
\Gamma _{\sigma ,\sigma ^{\prime }}^a({\bf k},{\bf k};i\nu _n,i\nu _n)$ and
similarly for $\gamma $] is found from a Ward identity. The standard Ward
identity (see e.g. Ref. \cite{MetznerRev})
\begin{equation}
{\rm lim}_{\omega \rightarrow 0}\Gamma _{\sigma \sigma ^{\prime }}^a({\bf k},%
{\bf k};i\nu _n,i\nu _n+i\omega )
\begin{array}{c}
=
\end{array}
\sigma _{\sigma \sigma ^{\prime }}^a[1-\partial \Sigma ({\bf k},i\nu
_n+i\omega )/\partial (i\omega )]_{\omega =0}
\end{equation}
is not appropriate for that purpose since it considers the opposite order of
limits for the vertex $\lim_{\omega \rightarrow 0}\lim_{q\rightarrow
0}\Gamma $ as required for the calculations in the RC regime. Instead, we
use the identity of Ref. \cite{HertzEdw} to obtain for $h\rightarrow 0$%
\begin{eqnarray}
\  &&\Gamma ^{+}({\bf k},{\bf k};i\nu _n,i\nu _n)=\Gamma ^z({\bf k},{\bf k}%
;i\nu _n,i\nu _n)=1+\left. \frac{\Sigma _{\uparrow }({\bf k},i\nu _n)-\Sigma
_{\downarrow }({\bf k},i\nu _n)}h\right| _{h=0}  \nonumber \\
\  &\simeq &1-\,\frac{TU}N\sum_{{\bf q},i\omega _n}G^2({\bf k}^{\prime
},i\nu _n^{\prime })\Gamma ({\bf k}^{\prime },{\bf k}^{\prime };i\nu
_n^{\prime },i\nu _n^{\prime })+\frac{TU^2}N\sum_{{\bf q},i\omega _n}\Big\{ %
\gamma ({\bf k},{\bf k+q};i\nu _n,i\nu _n+i\omega _n)  \nonumber \\
&&\ \ \ \times \Gamma ({\bf k+q},{\bf k+q};i\nu _n+i\omega _n,i\nu
_n+i\omega _n)G^2({\bf k}+{\bf q},i\omega _n+i\nu _n)[\chi ^{zz}({\bf q}%
,i\omega _n)-\chi ^{+-}({\bf q},i\omega _n)]  \nonumber \\
&&\ \ \ +\Big[{\frac d{dh}}\gamma ^z({\bf k},{\bf k+q};i\nu _n,i\nu
_n+i\omega _n)\Big]_{h=0}G({\bf k}+{\bf q},i\omega _n+i\nu _n)\chi ^{zz}(%
{\bf q},i\omega _n)\Big\}  \label{gg}
\end{eqnarray}
where we have denoted $\Gamma =\Gamma ^{+},$ $\gamma =\gamma ^{+}$ and we
have omitted spin indices. Note that in the third and fourth line of Eq. (%
\ref{gg} ) we have neglected $h$-derivatives of $\gamma ^{+},$ $\chi ^{+-},$
and $\chi ^{zz}.$ These derivatives lead to non-singular contributions, at
least for the linear dispersion case, which we mostly consider in the
following. Using the relation (\ref{Gg}) between $\Gamma $ and $\gamma $ and
the relation (\ref{hig}) between $\Gamma $ and $\chi ,$ we obtain from Eq. (%
\ref{gg}) the following integral equation for the vertex function $\gamma $
\begin{eqnarray}
&&\ \ \ \ \ \ \ \ \ \ \ \gamma ({\bf k},{\bf k};i\nu _n,i\nu _n)\,
\begin{array}{c}
\simeq
\end{array}
1+  \nonumber \\
&&\ \ \ \ \ \ \ \ \ \ \ \,\,\,\,\frac{TU^2}N\sum_{{\bf q},i\omega _n}\Big\{ %
\gamma ({\bf k},{\bf k+q};i\nu _n,i\nu _n+i\omega _n)\gamma ({\bf k+q},{\bf %
k+q};i\nu _n+i\omega _n,i\nu _n+i\omega _n)  \label{gg1} \\
&&\ \ \ \ \ \ \ \ \ \ \,\,\,\,\,\,\,\,\times G^2({\bf k}+{\bf q},i\omega
_n+i\nu _n)[\chi ^{zz}({\bf q},i\omega _n)-\chi ^{+-}({\bf q},i\omega _n)]
\nonumber \\
&&\ \ \ \ \ \ \ \ \ \ \,\,\,\,\,\,\,\,+{\frac{\chi ^{zz}({\bf q},i\omega _n)%
}{1+U\chi ({\bf q},i\omega _n)}}\Big[{\frac d{dh}}\gamma ^z({\bf k},{\bf k+q}%
;i\nu _n,i\nu _n+i\omega _n)\Big]_{h=0}G({\bf k}+{\bf q},i\omega _n+i\nu _n))%
\Big\}  \nonumber
\end{eqnarray}
In principle, Eq. (\ref{gg1}) is the first equation in an infinite
hierarchy, since the calculation of $d\gamma ^z/dh$ involves $d^2\gamma
^z/dh^2$ etc. To close the system of equations in an approximate way, we
proceed similarly to the $1/M$ expansion within the $O(M)$ generalization of
the Heisenberg model for localized-moment systems \cite{Chubukov,Quasi2D}.
Namely, we suppose that there are $M-1$ transverse spin modes and one
longitudinal mode. Then the correction due to $d\gamma ^z/dh$ in Eq. (\ref
{gg1}) is already of order $1/M$, and therefore only terms to leading order
in $1/M,$ arising from the transverse spin fluctuations can be retained when
calculating the derivative. To calculate $d\gamma ^z({\bf k},{\bf k};i\nu
_n,i\nu _n)/dh$ we differentiate the equation for the self-energy (\ref{Si}%
)\ twice with respect to the field. Note that the terms containing $%
d^2\gamma ^{\pm }/dh^2$ should be retained in this calculation, since they
are not small in $1/M$ and do not vanish in the $q\rightarrow 0$ limit
contrary to $d\gamma ^{\pm }/dh$. After lengthy algebraic manipulations we
finally obtain
\begin{eqnarray}
\  &&\ \ \ \Big[{\frac d{dh}}\gamma ^z({\bf k},{\bf k};i\nu _n,i\nu _n)\Big]%
_{h=0}=2[1+U\chi ({\bf 0},0)]  \nonumber \\
&&\ \ \ \ \ \ \times {\frac TN}\sum_{{\bf q},i\omega _n}\gamma ^3({\bf k},%
{\bf k+q};i\nu _n,i\nu _n+i\omega _n)G^3({\bf k}+{\bf q},i\omega _n+i\nu
_n)\chi ^{+-}({\bf q},i\omega _n)+{\cal O}(1/M)  \label{dg}
\end{eqnarray}

Generally, the susceptibilities (\ref{hifi}) which enter Eqs. (\ref{Si}) and
(\ref{gg}) can be represented in a form similar to Eq. (\ref{Hi1}) or (\ref
{Hi2}) at $\omega =0$,
\begin{equation}
\chi ^{+-}({\bf q},0)=2\chi ^{zz}({\bf q},0)\simeq \frac R{q^2+\xi ^{-2}}+%
\text{regular terms}  \label{Hih}
\end{equation}
where $R$ is a constant. As in localized-moment systems with strong
short-range order \cite{VIK}, the following identity
\begin{equation}
\langle {\bf S}_0\cdot{\bf S}_{{\bf r}}\rangle \simeq \frac TN\sum_{{\bf q}}
[\chi^{zz}({\bf q},0)+\frac{M-1}2\chi^{+-}({\bf q},0)]e^{i{\bf q\cdot r}}=
\overline{S}_0^2\exp (-r/\xi ),  \label{corr}
\end{equation}
which incorporates the correct long distance asymptotics of the real-space
correlation functions at $T\ll T^{*}$, is useful to relate $R$ and $%
\overline{S}_0$. As a result, we obtain
\begin{equation}
\overline{S}_0^2=\frac M{4\pi }TR\ln \xi .
\end{equation}

Since the dominant contribution to momentum sums in Eqs. (\ref{Si}) and (\ref
{gg}) arises from the long wavelength region $q\lesssim \xi ^{-1},$ we
replace
\begin{eqnarray}
\gamma _{\sigma ,\sigma ^{\prime }}^a({\bf k},{\bf k+q};i\nu _n,i\nu _n)
&\rightarrow &\gamma _{\sigma ,\sigma ^{\prime }}^a({\bf k},{\bf k};i\nu
_n,i\nu _n)\,,  \label{rep} \\
{\frac d{dh}}\gamma ^z({\bf k},{\bf k+q};i\nu _n,i\nu _n+i\omega _n)
&\rightarrow &{\frac d{dh}}\gamma ^z({\bf k},{\bf k};i\nu _n,i\nu _n)\,.
\nonumber
\end{eqnarray}
Equations (\ref{Si}), (\ref{gg}), and (\ref{dg}) together with the
replacements (\ref{rep}) form a closed system for $\gamma $ and $\Sigma .$
The remaining integrals over $q$ are easily calculated analytically
analogous to the non-self-consistent calculation in Section IIa. After the
continuation to the real axis we obtain the following algebraic equations
for the self-energy and the irreducible vertex $\gamma $ for $t\xi ^{-1}\ll
|\omega -\varepsilon _{{\bf k}}|\ll t,$
\begin{eqnarray}
\Sigma ({\bf k},\omega ) &=&\frac{\Delta _0^2\gamma ({\bf k},{\bf k};\omega
,\omega )}{\omega -\varepsilon _{{\bf k}}-\Sigma ({\bf k},\omega )}\,,
\nonumber \\
\gamma ({\bf k},{\bf k};\omega ,\omega ) &=&1-\frac{M-2}M\frac{\Delta
_0^2\gamma ^2({\bf k},{\bf k};\omega ,\omega )}{[\omega -\varepsilon _{{\bf k%
}}-\Sigma ({\bf k},\omega )]^2}+\frac 2M\frac{\Delta _0^4\gamma ^3({\bf k},%
{\bf k};\omega ,\omega )}{[\omega -\varepsilon _{{\bf k}}-\Sigma ({\bf k}%
,\omega )]^4}\,,
\end{eqnarray}
where we have introduced the ground state spin splitting $\Delta _0=U%
\overline{S}_0.$ Similar to the TPSC approach of Sect. IIa, the resulting
self-energy and the vertex depend on ${\bf k}$ only through $\overline{%
\omega }=\omega -\varepsilon _{{\bf k}}$ and are given by
\begin{eqnarray}
\Sigma ({\bf k},\omega ) &=&\frac{M\,(\Delta _0^2+\overline{\omega }^2-\sqrt{%
\overline{\omega }^2-\alpha _1\Delta _0^2}\,\sqrt{\overline{\omega }%
^2-\alpha _2\Delta _0^2}\,)}{2\,\left( 2+M\right) \,\overline{\omega }}
\nonumber \\
\gamma ({\bf k},{\bf k};\omega ,\omega ) &=&\frac M{2\,\left( 2+M\right)
^2\Delta _0^2\overline{\omega }^2}\,[2\overline{\omega }^4+(6\,+M)\Delta _0^2%
\overline{\omega }^2-M\Delta _0^4  \nonumber \\
&&+(M\Delta _0^2-2\overline{\omega }^2)\sqrt{\overline{\omega }^2-\alpha
_1\Delta _0^2}\,\sqrt{\overline{\omega }^2-\alpha _2\Delta _0^2}],
\label{SF}
\end{eqnarray}
where $\alpha _{1,2}=1+4(1\pm \sqrt{1+M/2})/M$; the branch Im$\sqrt{z}\geq 0$
of the square roots is chosen to guarantee the correct analytical properties
for $\Sigma $ and $\gamma .$ On the other hand, in the absence of the vertex
renormalization, i.e. for $\gamma ({\bf k},{\bf k};i\nu _n,i\nu _n)=1$,
which is the analogue of the FLEX approximation \cite{FLEX0} in our
approach, we obtain for the self-energy the $M$-independent result
\begin{equation}
\Sigma ({\bf k},\omega )=\frac 12(\overline{\omega }-\sqrt{\overline{\omega }%
-2\Delta _0}\sqrt{\overline{\omega }+2\Delta _0})\,.  \label{SSE}
\end{equation}

To zeroth order in $1/M$ (i.e. at $M=\infty $) we obtain from Eq. (\ref{SF})
\begin{equation}
\Sigma ({\bf k},\omega )
\begin{array}{c}
=
\end{array}
\frac{\Delta _0^2}{\overline{\omega }},\,\,\gamma ({\bf k},{\bf k};\omega
,\omega )
\begin{array}{c}
=
\end{array}
1-\frac{\Delta _0^2}{\overline{\omega }^2}  \label{NS}
\end{equation}
i.e. the self-energy and the vertex corrections exactly cancel each other at
$M=\infty $ and the self-energy and the vertex are given by their
second-order, non-self-consistent results. To demonstrate how this result
changes at finite values of $M$, we plot in Fig. 6 the results (\ref{SF})
and (\ref{SSE}) for $M=3$ together with the corresponding spectral
functions. Similarly to the non-self-consistent solution Eq. (\ref{NS}) the
real part of the self-energy in the self-consistent solution (\ref{SF}) is
divergent at $\omega =0.$ This is in contrast to the solution (\ref{SSE}),
which is self-consistent with respect to the self-energy only, and leads to
a finite positive slope of the real part of the self-energy at the Fermi
level, $\partial {\rm Re}\Sigma /\partial \omega =1/2$. The imaginary part
of the self-energy has its largest absolute value at the Fermi energy and
monotonically decreases with $|\omega |$ for the solution (\ref{SSE}), while
it has more complicated energy dependence with a $\delta $-function
singularity at $\omega =0$ according to the self-consistent result (\ref{SF}%
). As a consequence, the solution (\ref{SSE}), which is self-consistent with
respect to the self-energy only, displays a one-peak structure of the
spectral function, while Eq. (\ref{SF}) leads to a two-peak structure of the
spectral function (Fig. 6c), which is similar to the $M=\infty $ result (\ref
{NS}). We note that the gap in the spectral function at the Fermi energy in
the self-consistent solution is most likely an artifact of the first order
in $1/M$ result. At the same time the similarities between the results in
zeroth- and first order in $1/M$ allow to conclude that the higher order
corrections in $1/M$ are expected to be small and do not lead to qualitative
changes.

The one-peak structure of the spectral functions in the solution without
vertex renormalization is similar to that in the FLEX approximation in the
AFM case, which predicts only a slight variation of the quasiparticle weight
around the Fermi surface\cite{FLEX}. As discussed in Ref. \cite{TPSC}, this
result is a drawback of the absence of vertex corrections, and the necessity
to account for frequency-dependent vertex renormalization {\it together}
with the self-energy corrections is crucial in this case. As we have shown
in this section, a similar conclusion is reached in the FM case; the
corresponding non-trivial frequency dependence of the vertex is shown in
Fig. 6d.

Therefore, the solution (\ref{SF}) which accounts for both, self-energy and
vertex corrections leads to the results which are qualitatively similar to
those obtained in non-self-consistent approaches of Section II. The same
arguments can be applied to the $s$-$d$ model, which is therefore expected
to lead to the similar results at $T\ll T^{*}.$

\section{Summary and Conclusion}

We have investigated the self-energy in two dimensions in the vicinity of a
FM instability within the TPSC, the one-loop fRG analysis, and by applying
Ward identities. In all approaches the self-energy has a non-Fermi liquid
form in narrow window $|\omega |\lesssim \Delta _0$ around the Fermi level,
and the spectral functions have a two-peak structure at low temperatures.
The spectral weight at the Fermi energy is strongly suppressed at $T\lesssim
T^{*}$ in both, non-self-consistent and self-consistent solutions, because
the self-energy and the vertex corrections partially cancel each other. The
form of the spectral functions we have obtained in the paramagnetic state
with strong short-range magnetic order is qualitatively similar to the
mean-field result for the spectral functions in the FM ordered state, if the
electron spin quantization axis is chosen {\it perpendicular} to the
direction of the magnetization. Indeed, performing the simplest Stoner-like
decoupling of the interaction in the Hubbard model with the order parameter $%
\Delta =(U/2)\langle c_{i\uparrow }^{\dagger }c_{i\downarrow
}+c_{i\downarrow }^{\dagger }c_{i\uparrow }\rangle $, which corresponds to a
spin alignment along the $x$-axis, we readily obtain
\begin{equation}
\ll c_{{\bf k}\sigma }^{\dagger }\,|\,c_{{\bf k}\sigma }\gg _\omega =\frac 12%
\left( \frac 1{\omega -\varepsilon _{{\bf k}}-\Delta +i0^{+}}+\frac 1{\omega
-\varepsilon _{{\bf k}}+\Delta +i0^{+}}\right)   \label{G1}
\end{equation}
where $\ll |\,\gg _\omega $ denotes the Fourier transform of the retarded
Green function. With the inclusion of a small damping $i\gamma $ in the
denominators the Green function Eq. (\ref{G1}) leads indeed to a spectral
function, which is qualitatively similar to e.g. Fig. 6c. At the same time,
the mean-field Green functions for the spin projection {\it along} (parallel
or antiparallel) the magnetization axis are given by
\begin{equation}
\frac 12\ll c_{{\bf k}\uparrow }^{\dagger }\,\pm c_{{\bf k}\downarrow
}^{\dagger }|\,c_{{\bf k}\uparrow }\pm c_{{\bf k}\downarrow }\gg _\omega =%
\frac 1{\omega -\varepsilon _{{\bf k}}\mp \Delta +i0^{+}}\,.
\end{equation}
The spectral functions in the PM phase are necessarily spin-rotation
invariant. Therefore, we expect strong changes of the spectral functions at
the zero temperature FM phase transition only for a choice of the spin
quantization axis along the direction of the ground-state magnetization $%
{\bf M}$, while spectral functions for electrons with spin quantization axis
perpendicular to ${\bf M}$ change continuously with decreasing the
temperature to zero (or to the Curie temperature $T_C\ll t$ for quasi-2D
systems). Note also that beyond the mean-field approximation, the Green
functions for a spin projection antiparallel to ${\bf M}$ acquire strong
incoherent contributions, at least for fully polarized ferromagnets \cite
{HertzEdw,IK}. The overall qualitative picture for the evolution of the
spectral properties summarizes parts of the results of the present paper in
Fig. 7.

The two-peak structure of the spectral functions in the vicinity of a FM
instability leads to the formation of new coherent quasiparticles at the
points of the Brillouin zone near the spin-up and spin-down Fermi surfaces
of the FM ordered ground state. The spin-rotation symmetry remains however
necessarily unbroken at $T>0$ and new ``preformed'' qps do not have any
preferable spin direction. This $``$uncertainty'' of the spin direction in
2D magnets with strong short range order was earlier discussed in connection
with NMR experiments \cite{VIK1}, which are expected to give two-peak NMR
spectra, which is usually characteristic for 2-sublattice AFM order, even in
the vicinity of the FM state. For quasi-two-dimensional systems the
quasi-splitting of the Fermi surface is expected at $T_C<T\ll \Delta _0$
with $\Delta _0$ being the ground-state spin splitting.

Translated to real space, ``domains'' of size $\sim \xi $ may form
containing mostly electrons with a certain spin polarization$.$ Unlike in
localized-moments systems, however, these $``$domains'' have a small
electronic density and are therefore expected to be not a static but rather
a dynamic phenomenon. Such a dynamic formation of droplets with a preferred
spin direction near the FM instability should be distinguished from the
possible phase separation into hole-rich and hole-poor regions \cite
{Visher,Nagaev,Dagotto,Putikka} in the vicinity of an AFM instability for
the almost half-filled band. The latter may result as a compromise to
``adjust'' an AFM spin structure to an electron density $n<$1 and involve
charge fluctuations which are coupled to the spin channel for a nearly
half-filled band. At the same time, the dynamic domain formation near a FM
instability is expected solely due to FM spin fluctuations which favor the
aggregation of electrons with certain spin orientation in regions of size $%
\sim \xi .$

While the form of the spectral functions in the vicinity of the FM phase
transition is dramatically different from the conventional quasiparticle
one-peak structure, we do not expect sizable effects in the density of
states. Indeed, since the spectral functions depend at low-energy on $%
\varepsilon -\varepsilon _{{\bf k}}$ only, we find for $|\varepsilon |\ll t$
\begin{equation}
\rho (\varepsilon )={\frac 1N}\sum_{{\bf k}}A({\bf k},\varepsilon )\simeq
\int\limits_{\varepsilon _{\min }}^{\varepsilon _{\max }}{\rm d}\varepsilon
^{\prime }\rho _0(\varepsilon ^{\prime })A(\varepsilon -\varepsilon ^{\prime
})
\end{equation}
where $\rho _0(\varepsilon )$ is the bare density of states and $\varepsilon
_{\min ,\max }$ are the energies of the bottom and the top of the band with
respect to the Fermi level. Taking into account the two-peak structure of $%
A(\varepsilon )$ we obtain
\begin{equation}
\rho (\varepsilon )\simeq \frac 12[\rho _0(\varepsilon -\Delta )+\rho
_0(\varepsilon +\Delta )]
\end{equation}
i.e. the density of states is also ``pre-split'' at low $T$, but is not
strongly suppressed at the Fermi level.

Our results may provide a novel possibility to interpret ARPES data of
layered ferromagnetic materials. E.g., the described effects may be
important in the interpretation of the results of ARPES studies of the
layered manganite compound La$_{1+x}$Sr$_{2-x}$Mn$_2$O$_7$ \cite{Manganites}%
. Pseudogap structures are observed in this material both, above and below $%
T_C$. It was suggested that these structures are possibly produced by an
accompanying charge order\cite{Aliaga}, or by phase separation \cite{Moreo1}%
. The FM fluctuations however might be responsible for the part of the
pronounced shift ($\sim $250 meV) of the spectral weight maxima away from
the Fermi energy at the points of the Brillouin zone near the FM Fermi
surface with $T$ above the Curie temperature $T_C.$

The described effects may be also important for the normal state of some
unconventional superconductors, where ferromagnetic fluctuations are
expected to be important (c.f. UGe$_2,$ Sr$_2$RuO$_4$). UGe$_2$ has long
range magnetic order in the ground state and therefore may also show a
quasi-splitting of the FS above $T_C.$ Although long range FM order is
absent in strontium ruthenate Sr$_2$RuO$_4$, FM order is induced by a small
amount of La doping \cite{SrLa}.

Taking into account the strong self-energy and vertex renormalization in the
vicinity of the FM phase transition, the criteria for ferromagnetism \cite
{Moriya} should be reconsidered. This especially concerns quasi-2D systems,
where these renormalizations are expected to be the strongest. However, even
in 3D systems, where the finite-temperature divergencies of the self-energy
at the Fermi level are only logarithmic in $\xi $ (instead of a power law
divergence in 2D), they can also lead to additional renormalizations of the
Stoner criterion in comparison to those considered within the paramagnon
(spin fluctuation) theory of Refs. \cite{Moriya,Stamp}.

So far unresolved is the behavior of the self-energy in the quantum-critical
regime, if it exists at all for the FM phase transition due to the presence
of non-analytic contributions. This behavior depends crucially on the
temperature dependence of the correlation length; the $\varepsilon ^{2/3}$
dependence of the self-energy at $T=0$ however already implies a non-trivial
temperature dependence of the quasiparticle scattering rate at finite
temperatures. The simple ansatz for the magnetic susceptibility we have used
in the Ward identity approach is not justified in this case, and an
alternative approach has yet to be developed.

\newpage\ \centerline{\bf Acknowledgements}

We are grateful to W. Metzner for insightful discussions. This work was
supported by the Deutsche Forschungsgemeinschaft through SFB 484 and by
Grant No.~747.2003.2 (Support of Scientific Schools) from the Russian Basic
Research Foundation.

\newpage

\begin{figure}
\vspace{3cm}
\centerline{\psfig{file=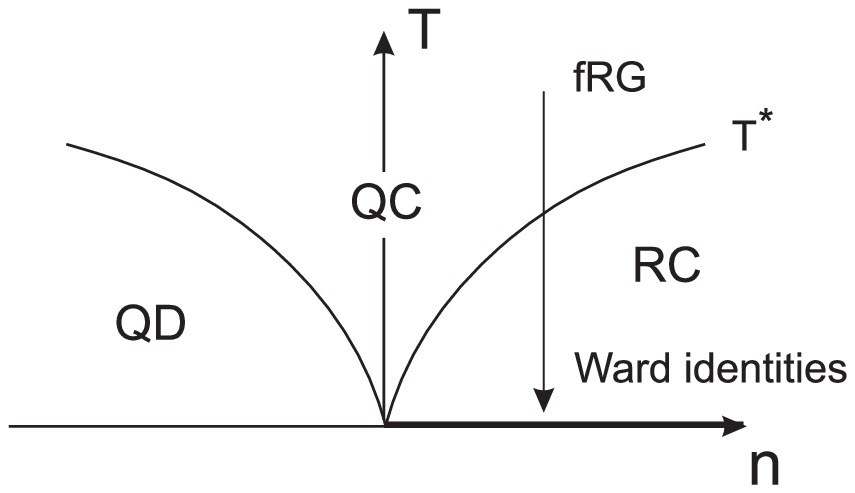}}
\vspace{2cm}
\caption{Schematic picture for the three different temperature regimes, as
proposed in Ref. [25]: (1) the renormalized classical (RC) regime
above the ordered ground state (indicated by the bold line, $T^{*}$ is the
corresponding crossover temperature scale discussed in the text), (2) the
quantum disordered (QD) regime above a disordered ground state, and (3) the
quantum critical (QC) regime. The arrow shows the direction, in which the
evolution of the spectral properties is traced in this paper; ``fRG'' and
``Ward identities'' in the figure mark the intermediate ($T\gtrsim T^{*}$)
and low temperature ($T\ll T^{*}$) regimes, where the corresponding
approaches are applied. The two-particle self-consistent approach is applied
in both regimes.}
\end{figure}

\newpage

\begin{figure}
\psfig{file=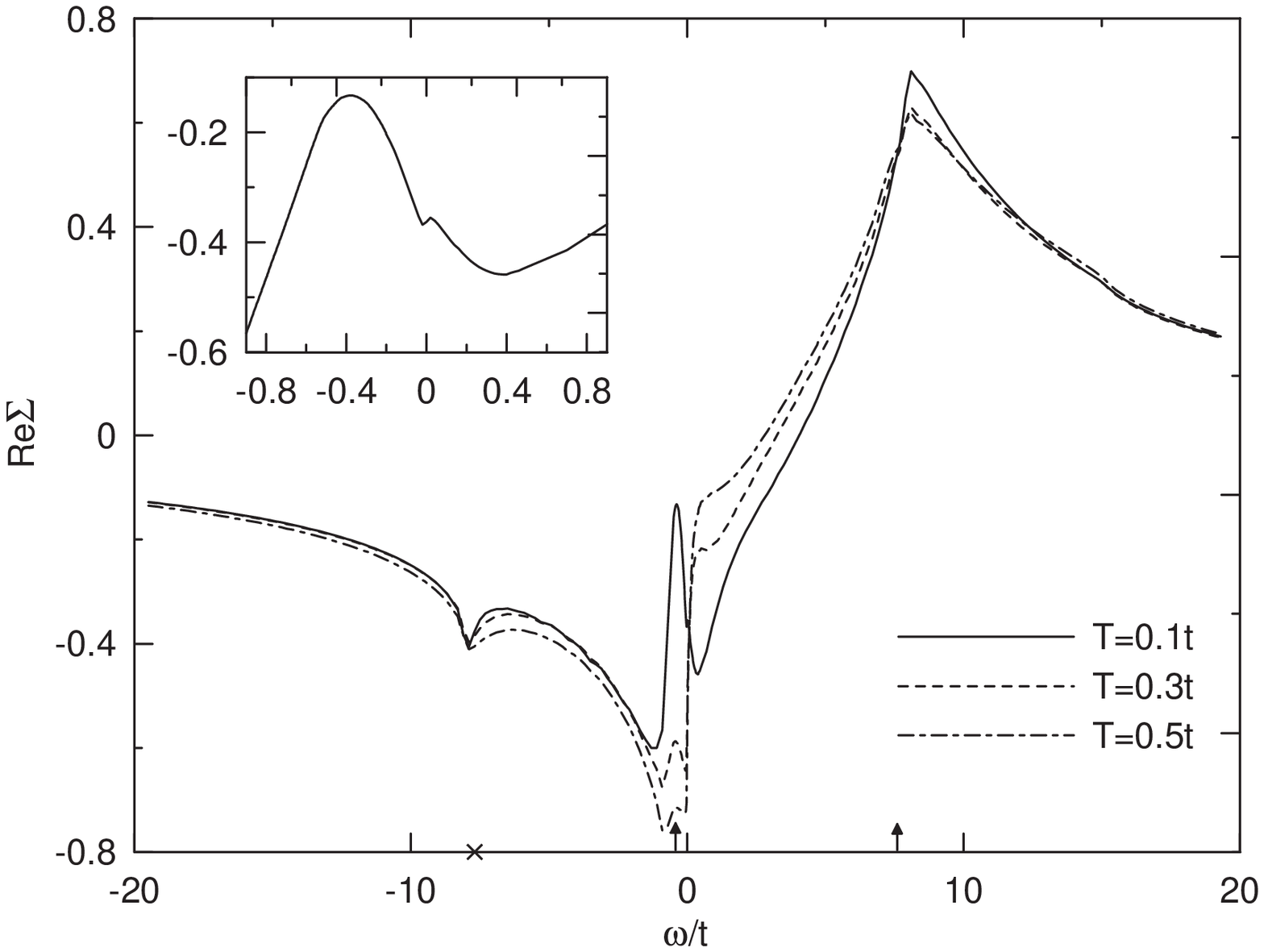}
\vspace{1cm}
\psfig{file=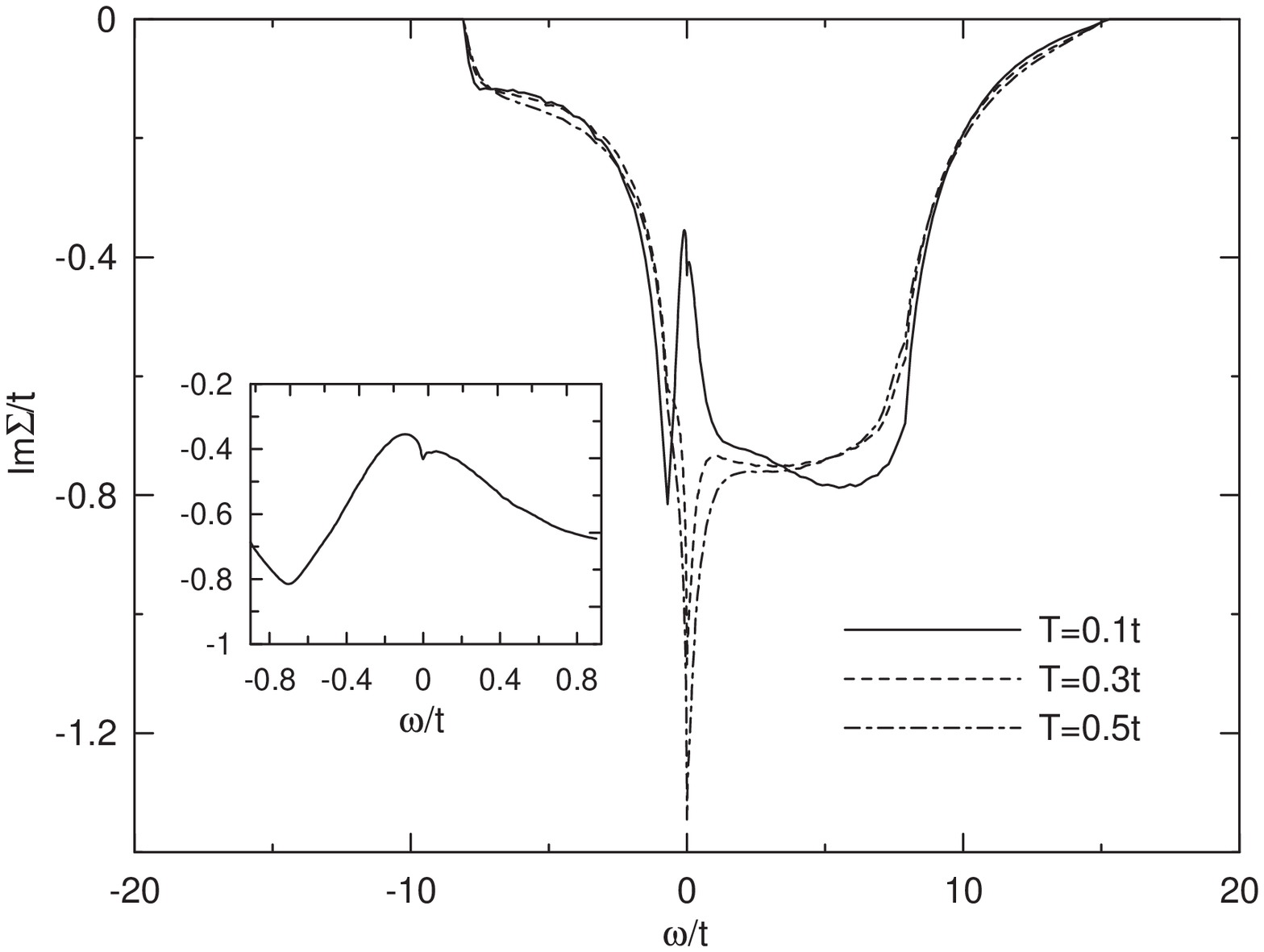}
\vspace{1cm}
\caption{The real (a) and imaginary (b) parts of the self-energy in
second-order perturbation theory (SOPT) at $t^{\prime }/t=0.45,$ $U=4t,$ $%
\mu =0$ (vH band filling), and different temperatures. Arrows mark the
energy of the lower- ($\varepsilon _{\min }=-4+8t^{\prime }$) and upper- ($%
\varepsilon _{\max }=4+8t^{\prime }$) edges of the noninteracting band, the
cross marks the energy, opposite to the upper edge of the band, $%
-\varepsilon _{\max }.$ The inset shows the behavior at low energies for $%
T=0.1t$.}
\end{figure}

\newpage

\begin{figure}
\psfig{file=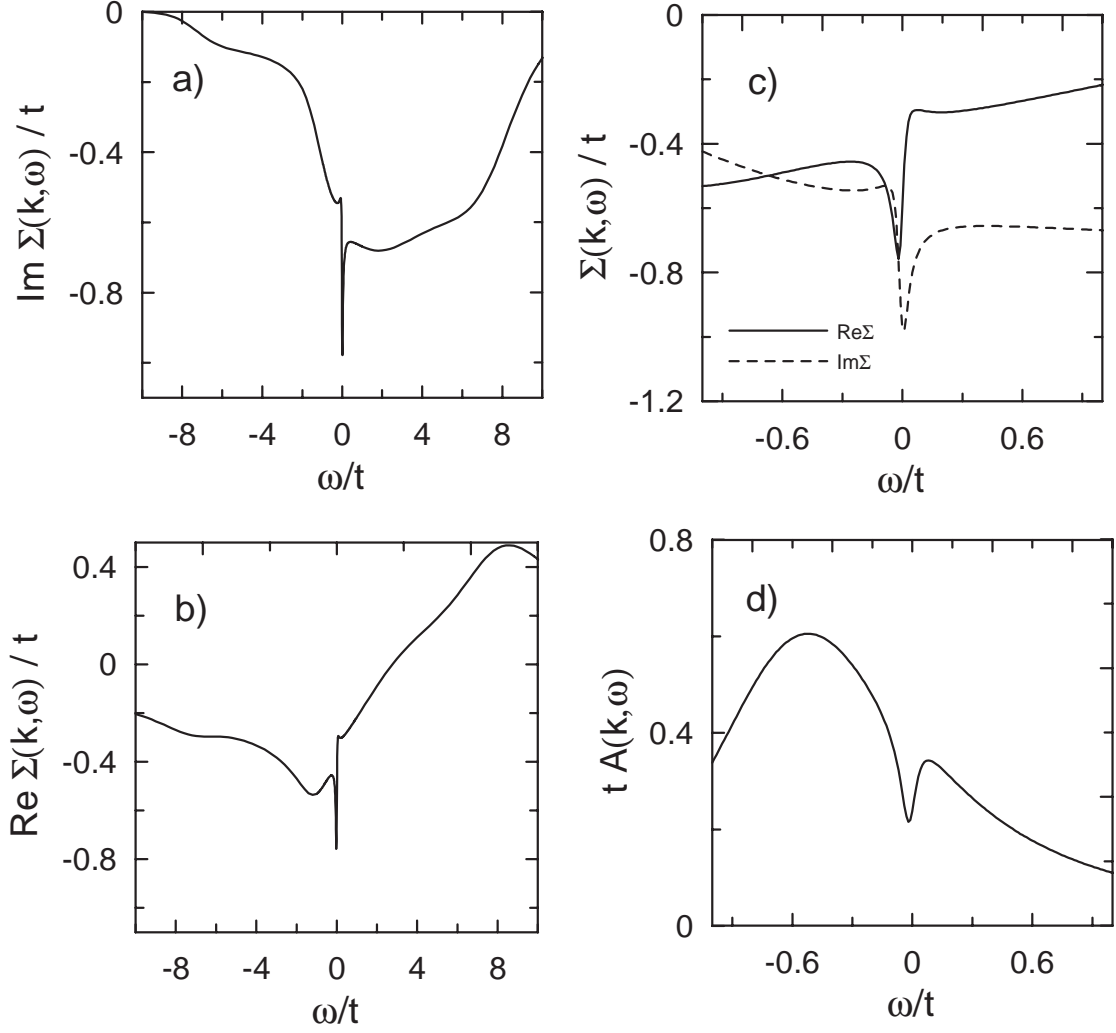}
\vspace{1cm}
\caption{The functional renormalization group results for the self-energy
(a-c) and the spectral function (d) in the first Fermi surface patch,
closest to the ($\pi ,0$) point at $t^{\prime }/t=0.45,$ $U=4t,$ $\mu =0,$
and $T=0.3t$.}
\end{figure}

\newpage

\begin{figure}
\psfig{file=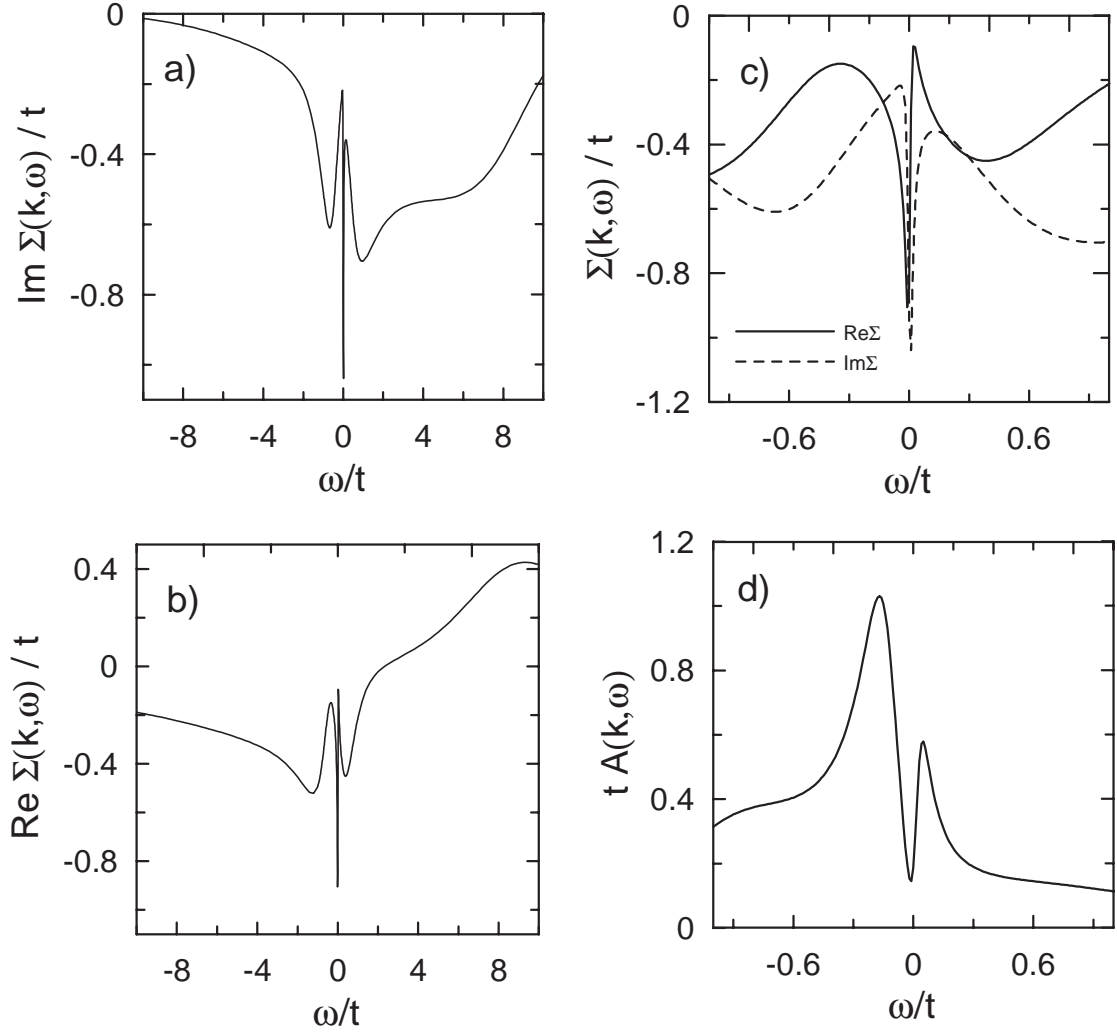}
\vspace{1cm}
\caption{Same as Fig. 3 for $T=0.1t.$}
\end{figure}
\newpage

\begin{figure}
\psfig{file=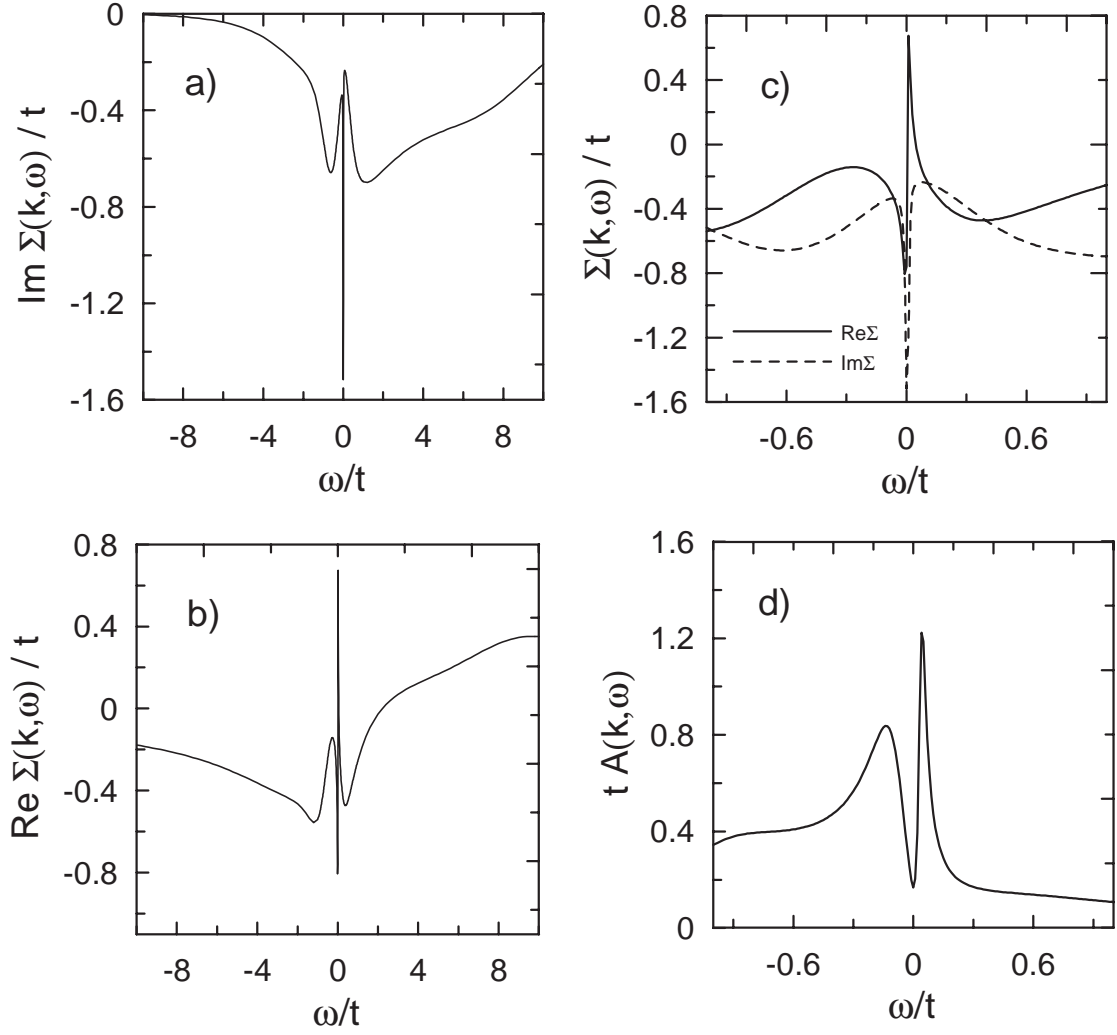}
\vspace{1cm}
\caption{Same as Fig. 3 for $T=0.1t$ in the fourth Fermi surface patch
closest to the Brillouin zone diagonal.}
\end{figure}

\newpage

\begin{figure}
\psfig{file=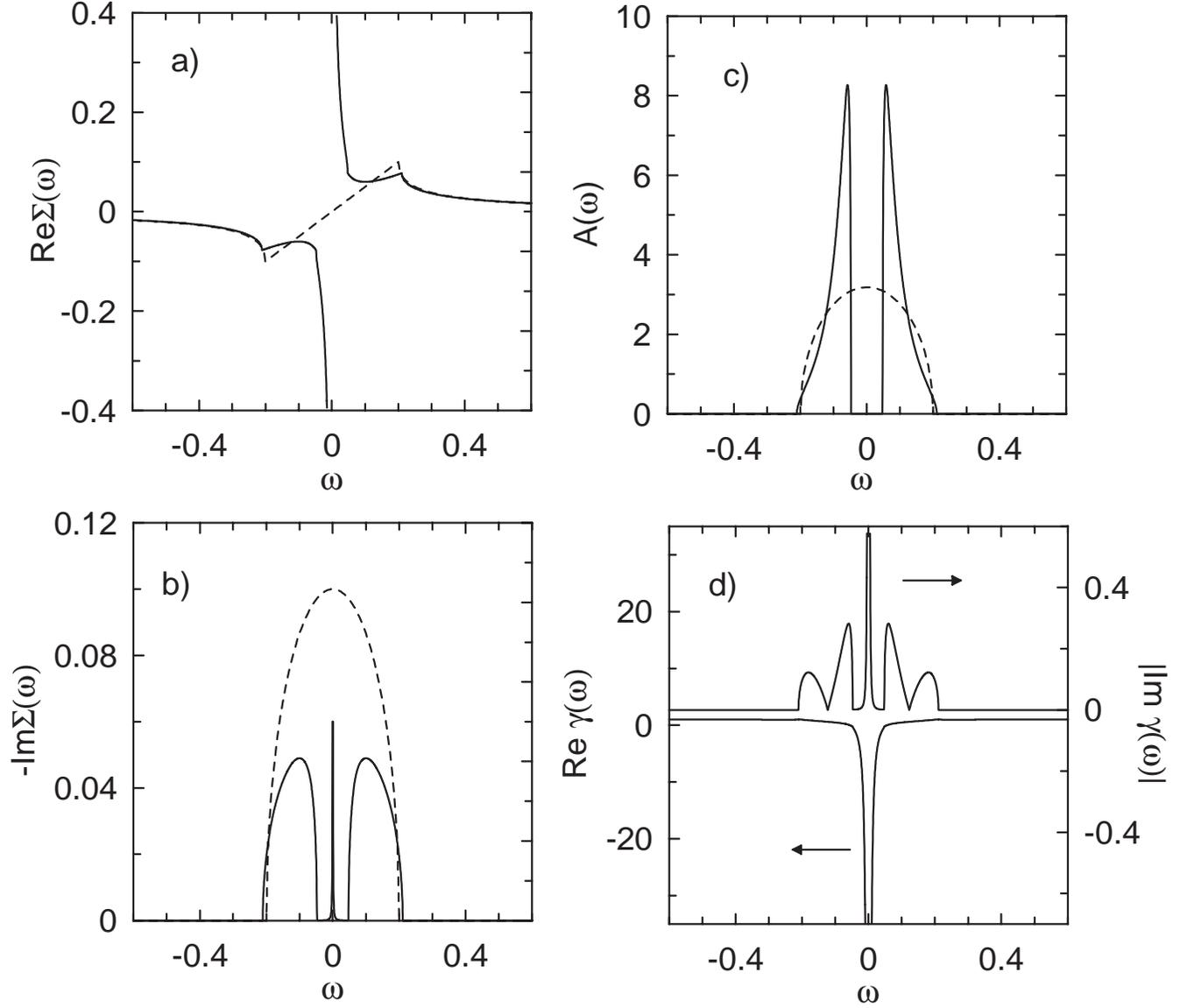}
\vspace{1cm}
\caption{The real and imaginary parts of the self-energy (a,b), the spectral
function (c), and the vertex function $\gamma $ (d) at $T\ll T^{*}$ in the
fully self-consistent Ward identity approach (solid lines) and the FLEX-like
approach, which is self-consistent with respect to the self-energy only
(dashed lines) as a function of $\overline{\omega }=\omega -\varepsilon _{%
{\bf k}}$.}
\end{figure}

\newpage

\begin{figure}
\centerline{\psfig{file=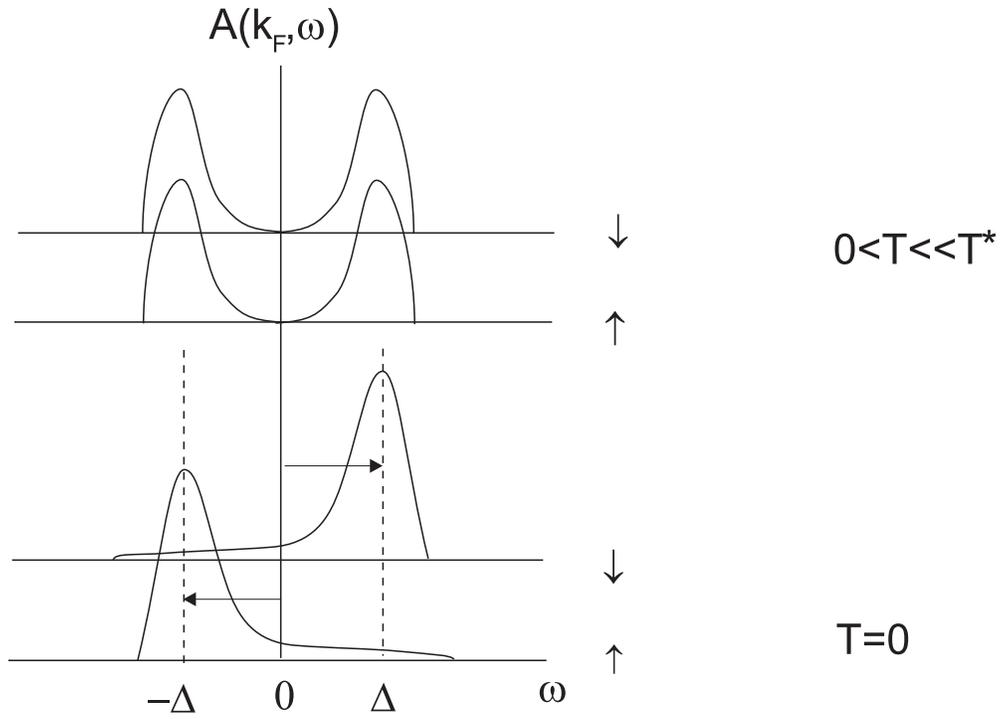}}
\vspace{1cm}
\caption{Schematic picture for the evolution of the spectral function near
the ferromagnetic phase transition. The spectral functions at $T=0$ are
shown at the paramagnetic Fermi surface. The spectral functions at the Fermi
surfaces of spin-up and spin-down electrons are expected to shift by $\pm
\Delta$ with respect to those of the paramagnetic Fermi surface, as
indicated by arrows.}
\end{figure}

\end{document}